\documentclass[aps,pra,twocolumn,groupedaddress,showpacs]{revtex4}
\usepackage{epsf,epsfig,graphics,graphicx}
\begin{document}

\title{(Quasi)-exactly solvable quasinormal modes}

\author{Hing-Tong Cho}
  \email{htcho@mail.tku.edu.tw}
\author{Choon-Lin Ho}
  \email{hcl@mail.tku.edu.tw}
\affiliation{Department of Physics, Tamkang University, Tamsui
251, Taiwan, Republic of China}

%\date{Dec 8, 2006}

\begin{abstract}
We consider quasinormal modes with complex energies from the point
of view of the theory of quasi-exactly solvable (QES) models. We
demonstrate that it is possible to find new potentials which admit
exactly solvable or QES quasinormal modes by suitable
complexification of parameters defining the QES potentials.
Particularly, we obtain one QES and four exactly solvable
potentials out of the five one-dimensional QES systems based on
the $sl(2)$ algebra.
\end{abstract}

 \maketitle

{\it Introduction.}~Quasinormal modes (QNM) arise as perturbations
of stellar or black hole spacetimes \cite{QNM}. They are solutions
of the perturbation equations that are outgoing to spatial
infinity and the event horizon.  Such ``outgoing wave boundary
condition" was first adopted by Gamow in his explanation of the
$\alpha$-decay of atoms as a quantum tunneling process
\cite{alpha}. Generally, these conditions lead to a set of
discrete complex eigenfrequencies, with the real part representing
the actual frequency of oscillation and the imaginary part
representing the damping.  QNM carry information of black holes
and neutron stars, and thus are of importance to
gravitational-wave astronomy.  In fact, these oscillations,
produced mainly during the formation phase of the compact stellar
objects, can be strong enough to be detected by several large
gravitational wave detectors under construction.  Recently, QNM of
particles with different spins in black hole spacetimes have also
received much attention \cite{spin}.

Owing to the intrinsic complexity in solving the perturbation
equations in general relativity with the appropriate boundary
conditions, one has to resort to various approximation methods,
eg., the WKB method, the phase-integral method etc., in obtaining
QNM solutions.  It is therefore helpful that one can get some
insights from exact solutions in simple models, such as the
inverted harmonic oscillator \cite{Barton,Kim} and the
P\"oschl-Teller potential \cite{FM}.  Unfortunately, the number of
exactly solvable models is rather limited.

Recently, in non-relativistic quantum mechanics a new class of
potentials which are intermediate to exactly solvable ones and
non-solvable ones has been found. These are called quasi-exactly
solvable (QES) problems for which it is possible to determine
analytically a part of the spectrum but not the whole spectrum
\cite{TU,Tur,GKO,Ush,PT,KMO}.  The discovery of this class of
spectral problems has greatly enlarged the number of physical
systems which we can study analytically.  In the last few year,
QES theory has also been extended to the Pauli and Dirac equations
\cite{Rel}.

In this paper we would like to study solutions of QNM based on the
Lie-algebraic approach of QES theory.  We demonstrate that, by
suitable complexification of some parameters of the generators of
the $sl(2)$ algebra while keeping the Hamiltonian Hermitian, we
can indeed obtain potentials admitting exact or quasi-exact QNMs.
Such consideration has not been attempted before in studies of QES
theory.  Our work represents a direct opposite of the work in
\cite{BB}, where QES {\it real energies} were obtained from a {\it
non-Hermitian} {\cal PT}-symmetric  Hamiltonian.

{\it QES Theory.}~Let us briefly review the essence of the
Lie-algebraic approach \cite{TU,Tur,GKO} to QES models
\cite{other}. Consider a Schr\"odinger equation $H\psi=E\psi$ with
Hamiltonian $H=-d_x^2 + V(x)$ ($d_x\equiv d/dx$) and wave function
$\psi (x)$. Here $x$ belongs either to the interval
$(-\infty,\infty)$ or $[0,\infty)$. Now suppose we make an
``imaginary gauge transformation" on the function $\psi$: $\psi
(x)= \chi(x) e^{-g(x)}$, where $g(x)$ is called the gauge
function. For physical systems which we are interested in, the
phase factor $\exp(-g(x))$ is responsible for the asymptotic
behaviors of the wave function so as to ensure normalizability.
The function $\chi(x)$ satisfies a Schr\"odinger equation with a
gauge transformed Hamiltonian $H_g=e^g H e^{-g}$. Suppose $H_g$
can be written as a quadratic combination of the generators $J^a$
of some Lie algebra with a finite dimensional representation.
Within this finite dimensional Hilbert space the Hamiltonian $H_g$
can be diagonalized, and therefore a finite number of eigenstates
are solvable. Then the system described by $H$ is QES.  For
one-dimensional QES systems the most general Lie algebra is
$sl(2)$, and $H_g$ can be expressed as
\begin{eqnarray}
H_g=\sum C_{ab}J^a J^b + \sum C_a J^a + {\rm real\ constant}~,
\label{H-g}
\end{eqnarray}
 where $C_{ab},~C_a$ are taken to be \emph{real constants} in  \cite{Tur,GKO}.
The generators $J^a$ of the $sl(2)$ Lie algebra  take the
differential forms: $ J^+ = z^2 d_z - nz~,~ J^0=z
d_z-n/2~,~J^-=d_z$ ($n=0,1,2,\ldots$). The variables $x$ and $z$
are related by some function to be described later. $n$ is the
degree of the eigenfunctions $\chi$, which are polynomials in a
$(n+1)$-dimensional Hilbert space with the basis $\langle
1,z,z^2,\ldots,z^n\rangle$.

Substituting the differential forms of $J^a$ into Eq.~(\ref{H-g}),
one sees that every QES operator $H_g$ can be written in the
canonical form : $H_g=-P_4(z) d_z^2 + P_3(z)d_z + P_2(z)$, where
$P_k (z)$ are $k$-th degree polynomial in $z$ with real
coefficients related to the constants $C_{ab}$ and $C_a$. The
relation between $H_g$ and the standard Schr\"odinger operator $H$
fixes the required form of the gauge function $g$ and the
transformation between the variable $x$ and $z$. Particularly,
$x=\int^z dy/\sqrt{P_4 (y)}$.  Analysis on the inequivalent forms
of real quartic polynomials $P_4$ thus give a classification of
all $sl(2)$-based QES Hamiltonians \cite{Tur,GKO}.  If one imposes
the requirement of non-periodic potentials, then there are only
five inequivalent classes, which are called case 1 to 5 in
\cite{GKO}.

Our main observation is this.  If some of the coefficients in $P_k
(z)$ are allowed to be complex while keeping $V(x)$ real, then all
the five cases classified in  \cite{GKO} can indeed support
QES/exact quasinormal modes.  We shall discuss these cases below.

{\it{QES QNM.}}~We consider Case 3 in \cite{GKO}, which
corresponds to Class I potential in Turbiner's scheme \cite{Tur}.
There are two subclasses in this case, namely, Case (3a) and (3b).
We shall present the analysis of QNM potential for case (3a). The
other case turns out to give the same potential with a suitable
choice of the parameters. The potential in case (3a) has the form
(in this paper we adopt the notation of \cite{GKO}):
\begin{eqnarray}
V(x)=A e^{2\sqrt{\nu} x}+ B e^{\sqrt{\nu} x}+ C e^{-\sqrt{\nu} x}
+ D e^{-2\sqrt{\nu} x}~, \label{V-3a}
\end{eqnarray}
where $x\in (-\infty,\infty)$ and $\nu$ is a positive scale
factor. Note that $V(x)$ is defined up to a real constant, which
we omit for simplicity, as it merely shifts the real part of the
energy. This remark also applies to the other cases.   $V(x)$ in
Eq.~(\ref{V-3a}) reduces to the exactly solvable Morse potentials
when $A=B=0$, or $C=D=0$. This potential is QES when the
coefficients are related by
\begin{eqnarray}
 A&=& \frac{\hat{b}^2}{4\nu}~,~~~
 B=\frac{\hat{c}+(n+1)\nu}{2\nu}~\hat{b},~~~\nonumber\\
 C&=&\frac{\hat{c}-(n+1)\nu}{2\nu}~\hat{d},~~~
 D=\frac{\hat{d}^2}{4\nu},~\\
 &&~n=0,1,2\ldots\nonumber
 \end{eqnarray}
Here $\hat{b},~\hat{c},~\hat{d}$ are arbitrary real constants. For
each integer $n\geq 0$, there are $n+1$ exactly solvable
eigenfunctions in the $(n+1)$-dimensional QES subspace:
\begin{equation}
\psi_n(x)=\exp{\left[\frac{\hat{b}}{2\nu} e^{\sqrt{\nu} x} +
\frac{\hat{c}-n\nu}{2\sqrt{\nu}} x - \frac{\hat{d}}{2\nu}
e^{-\sqrt{\nu} x}\right]}\chi_n (e^{\sqrt{\nu} x})~. \label{psi-n}
\end{equation}
Here $\chi_n (z)$ is a polynomial of degree $n$  in
$z=\exp(\sqrt{\nu}x)$.  To guarantee normalizability of the
eigenfunctions, the real constants
$\hat{b},~\hat{c},~\hat{d},~\nu$ and $n$ must satisfy certain
relations \cite{GKO}.

\begin{figure}[!]
\includegraphics{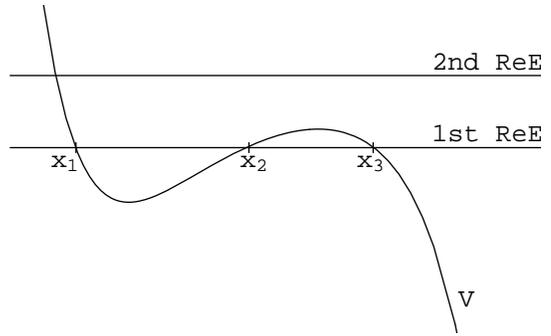}
\caption{\label{fig1} Schematic sketch of the potential $V(x)$ in
Eq.~(\ref{V-3a1}) and the real parts of the corresponding QNMs
with $b=1$ and $n=1$.}
\end{figure}

We want to see if we can get QNM solutions if we allow some
parameters to be complex, while still keeping the potential $V(x)$
real. This latter requirement restricts the possible values of the
parameters, and hence the forms of QES potential admitting
quasinormal modes. For the case at hand, we find that one possible
choice of values of $\hat{b},~\hat{c}$ and $\hat{d}$ is:
\begin{equation}
\hat{b}=ib ,~\hat{c}=-(n+1)\nu,~\hat{d}=d,~~b,d:{\rm real\
constants}~.
\end{equation}
The potential Eq.~(\ref{V-3a}) becomes
\begin{eqnarray}
V_n(x)=-\frac{b^2}{4\nu} e^{2\sqrt{\nu} x}- \left(n+1\right) d
e^{-\sqrt{\nu} x} + \frac{d^2}{4\nu} e^{-2\sqrt{\nu} x}~,
\label{V-3a1}
\end{eqnarray}
and the wave function Eq.~(\ref{psi-n}) becomes
\begin{eqnarray}
\psi_n(x)&=&\exp{\left[\frac{ib}{2\nu} e^{\sqrt{\nu} x} -
\left(n+\frac{1}{2}\right)\sqrt{\nu} x - \frac{d}{2\nu}
e^{-\sqrt{\nu} x}\right]}\nonumber\\ &&~~~~~\times \chi_n
(e^{\sqrt{\nu} x})~. \label{psi-n1}
\end{eqnarray}
$V(x)$ approaches $\mp\infty$ as $x\to\pm\infty$ respectively: it
is unbounded from below on the right. For small positive $d$ and
sufficiently large $n$, $V(x)$ can have a local minimum and a
local maximum. In this case, the well gets shallower as $d$
increases at fixed value of $n$, or as $n$ decreases at fixed $d$.
Fig.~\ref{fig1} presents a schematic sketch of $V(x)$ with $b=1$
and $n=1$. We emphasize here that for different value of $n$, each
$V(x)$ represents a different QES potential admitting $n+1$ QES
solutions. Since $V(x)\to \infty$ as $x\to -\infty$, the wave
function must vanish in this limit. This means $d>0$ from
Eq.~(\ref{psi-n1}). For the outgoing boundary condition, we must
take $b>0$.  Before we go on, we note here that there is another
possible choice of the parameters, namely,
\begin{equation}
\hat{b}=-b ,~\hat{c}=(n+1)\nu,~\hat{d}=id,~~b,d:{\rm real\
constants}~.
\end{equation}
However, this choice leads to a potential related to
Eq.~(\ref{V-3a1}) by the reflection $x\mapsto -x$.  Hence, we will
only discuss the potential in Eq.~(\ref{V-3a1}) here.

To see that the wave functions $\psi_n(x)$ do represent
quasinormal modes, we determine the corresponding energy $E_n$.
This is easily done by solving the eigenvalue problem of the
polynomial part $\chi_n (z)$ of the wave function.  From the
Schr\"odinger equation we find that for $n=0$, the energy is $E_0=
-\nu/4- ibd/2\nu$.  This clearly shows that the only QES solution
when $n=0$ is a QNM with an energy having a negative imaginary
part (recall that $b,d,\nu>0$). For $n=1$, we have two QES
solutions. Their energies are $E_1=-5\nu/4- ibd/2\nu \pm
\sqrt{\nu^2 -ibd}$. Again we have two QNM modes. One can proceed
accordingly to obtain $n+1$ QNM modes with higher values of $n$.
However, for large $n$, computation becomes tedious, and one has
to resort to numerical means.  For definiteness, we list in
Table~\ref{table1} some values of $E_n$ for the case where
$b=d=\nu=1$.  In this case, the potential has a local minimum and
a local maximum, with the barrier height rises as $n$ increases.
For large $n$, some states (in parentheses)
 have their real parts of energy lie below the local
maximum of the barrier . Such states are conventionally called the
metastable states. We see from the table that the number of
metastable states contained within the well increases as $n$
becomes larger.  This is reasonable as the well becomes deeper as
$n$ increases.

We have also used the WKB method to determine possible metastable
states that can be trapped by the well.  The method is as follows.
Let the three turning points, from left to right, be designated as
$x_1,x_2$ and $x_3$, as shown in Fig~\ref{fig1}. The boundary
conditions of QNM then leads to the following quantization
condition for the energy $E$:
\begin{eqnarray}
e^{2i\beta}=\frac{1+4e^{2\gamma}}{1-4e^{2\gamma}}~, \label{WKB}
\end{eqnarray}
where $\beta \equiv \int_{x_1}^{x_2} dx~\sqrt{E-V(x)}$, and
$\gamma \equiv  \int_{x_2}^{x_3} dx~\sqrt{V(x)-E}$.

\begin{table}[!]
 \caption{\label{table1}Values of QNM energy
$E_n$ for the QES potentials  in case 3 with parameters
$b=d=\nu=1$. Note that for each $n$, there are $n+1$ values of
$E_n$. Energy levels of metastable states are in parentheses. WKB
estimates of energy of metastable states are also listed.}
\begin{ruledtabular}
\begin{tabular}{ccccc}
%&& &{$E_n$}&\\ \cline{2-5}
 n &&{$E_n$ (QES)}  &&{$E_n$ (WKB)}\\
 \hline \hline
 0 && $-0.25-0.5$i && --- \\ \hline
 1 && ($-2.349- 0.0449$i) && $-2.313 -0.0588$i\\
   && $-0.151-0.955$i && --- \\ \hline
 2 && ($-6.271-0.000140$i) && $-6.267-0.000147$i\\
   && ($-2.447 -0.135$i) &&  $-2.439- 0.156$i\\
   && $-0.0317 -1.365$i && --- \\ \hline
 3 && ($-12.258 -4.786\times 10^{-8}$i) && $-12.257 -4.813\times 10^{-8}$i\\
   && ($-6.293-0.000817$i) && $-6.284 - 0.000902$i \\
   && $-2.542-0.259$i && --- \\
   && $0.0939 - 1.740$i && --- \\ \hline
 4 && ($-20.255- 4.310\times 10^{-12}$i) && $-20.254 -4.328\times 10^{-12}$i\\
   && ($-12.265 - 3.810\times 10^{-7}$i) && $-12.263-4.010\times 10^{-7}$i \\
   && ($-6.323-0.002742$i) && $-6.306-0.00311$i\\
   && $-2.628 - 0.406$i && --- \\
   && $0.220 - 2.091$i && --- \\
\end{tabular}
\end{ruledtabular}
\end{table}

To the first approximation, we set $e^{-2\gamma}\approx 0$ and
$E\approx Re(E)$.  This gives the quantization condition of
$Re(E)$: $\beta=(l+1/2)\pi~,~l=0,1,2\ldots$. Then by keeping the
imaginary part of $E$ in $\beta$, but dropping it in $\gamma$, we
can obtain an estimate of $Im(E)$:
\begin{eqnarray}
Im(E)= -\frac{e^{-2\int_{x_2}^{x_3} dx \sqrt{V(x)-Re(
E)}}}{2\int_{x_1}^{x_2} ~dx/\sqrt{Re(E)-V(x)}}~.
\end{eqnarray}
In Table~\ref{table1} we have also listed the WKB results for the
metastable states.  It is interesting to note that all the
metastable states obtained by WKB methods are in fact the QES
states in the cases we considered.  The exact values of the QES
energies and those of WKB calculations are seen to be consistent.

{\it Oscillator-like potentials.}~We now turn to the other four
cases. These four cases admit \emph{exact} QNM solutions. Here we
shall discuss Case 4 and 5 of \cite{GKO}, which are associated
with oscillator potentials.

For Case 4, the potential is given by
\begin{eqnarray}
V(x)=A x^6 + B x^4 + C x^2 + \frac{D}{x^2}~,~~~ x\in [0,\infty)~.
\end{eqnarray}
It is QES if the coefficients are related by
\begin{eqnarray}
A&=&\frac{\hat{b}^2}{256}~,~~
B=\frac{\hat{b}\hat{c}}{32}~,~~\nonumber\\
C&=&\frac{1}{16}\left[\hat{c}^2 +
\left(2\hat{d}+3(n+1)\right)\hat{b}\right]~,\nonumber\\
~~D&=&\left(\hat{d}-\frac{n}{2}\right)\left(\hat{d}-\frac{n}{2}
-1\right)~, ~~n=0,1,\ldots
\end{eqnarray}
with real constants $\hat{b}~,\hat{c}$ and $\hat{d}$. As before,
we now relax the reality constraint on the parameters but keeping
$V(x)$ real, and determine if QNMs can be supported in this case.
It turns out the answer is positive, if we let $\hat{b}=0$,
$\hat{c}=4ia$, and $\hat{d}=d$, with real $a$ and $d$.  This leads
to the potential
\begin{eqnarray}
V(x)=-a^2 x^2 + \frac{\gamma\left(\gamma -1\right)}{x^2}~.
\label{V-4}
\end{eqnarray}
Here $\gamma\equiv d-\frac{n}{2}$ is arbitrary. Class VII and VIII
in \cite{Tur} also lead to this potential when some parameters are
allowed to be complex. Eq.~(\ref{V-4}) is independent of $n$, and
hence one can solve for solvable states with any degree $n$ in its
polynomial part. This system is therefore exactly solvable. For
$\gamma >1$ and $\gamma <0$, $V(x)$ is a monotonic decreasing
function on the positive half-line.  If $0<\gamma <1$, $V(x)\to
-\infty$ at $x=0,\infty$, and has a global maximum in between.

The wave functions take the form $\psi_n=x^\gamma e^{ia
x^2/2}~\chi_n(x^2)$, where $\chi_n(x^2)$ is an $n$-th degree
polynomial in $z\equiv x^2$.  In order that the wave function
satisfies the outgoing wave condition and vanishes at the origin,
we must have $a>0,~\gamma
>0$.  Plucking $\psi_n(x)$ into the Schr\"odinger equation, we
determine the energies to be $E_n=-i(4n + 2\gamma +1)a$. Since $a$
and $\gamma$ are positive, $E_n$ is always negative, indicating
decaying QNMs.

Case 3 and 4 illustrate the main steps in complexifying the
relevant parameters to obtain QES/exact QNMs.  It is satisfying to
find that the remaining three cases can also be extended to give
potentials which also admit exact QNMs.  These cases will be
treated briefly below.

For Case 5 we find that the only viable choice of potential is
$V(x)=-(cx+d)^2 + d^2/4$, where $c$ and $d$ are real constants.
This is a shifted inverted oscillator.  Just as in Case 4, here
$V(x)$ is also independent of $n$, and hence also exactly
solvable. We mention that Class VI in \cite{Tur} also leads to
this potential with appropriate choice of parameters.

For simplicity, we briefly discuss the case with $d=0$: $V(x)=-c^2
x^2/4$ \cite{Barton}.  QNM solutions in such inverted oscillator
has been discussed in \cite{Kim} using the modified annihilation
and creation operators.  Here we consider the problem from the
point of QES potential. The wave function is given by $\psi_n (x)
=\exp(icx^2/4)\chi_n (x)$.  From the Schr\"odinger equation, we
get the energies $E_n=-ic(n + 1/2)$.  For $c>0$, the wave function
describes decaying outgoing QNM away from the maximum $x=0$ to
$x=\pm\infty$. This is analogous to the QNM in black holes.  When
$c<0$, we have growing incoming QNM moving towards the origin.
This latter case was obtained in \cite{Kim}.

{\it Hyperbolic potentials.}~Potentials in Case 1 and 2 involve
hyperbolic functions. A proper complexification of the parameters
in Case 1 is
\begin{eqnarray}
V(x)&=&-\frac{cd}{2\nu}\tanh(\sqrt{\nu} x){\rm sech}(\sqrt{\nu}
x)\nonumber\\ & +&\frac{1}{4\nu}\left(\nu ^2 +c^2 - d^2\right){\rm
sech}^2(\sqrt{\nu} x)~, \label{V1}
\end{eqnarray}
where $x\in (-\infty,\infty)$. Its normalizable counterpart is,
according to the classification in \cite{Cooper}, the exactly
solvable Scarf II potential. The special case where $d=0$, which
is the inverted P\"oschl-Teller potential, has been employed in
\cite{FM} in their study of black hole's QNMs.  We can easily
obtain the wave functions and energies for the general case,
Eq.~(\ref{V1}), from QES theory. The wave function has the form:
  $\psi_n(x)=(\cosh\sqrt{\nu} x)^{(ic+\nu)/2\nu}\exp(id
\tan^{-1}(\sinh\sqrt{\nu}x)/2\nu)\times\chi_n(\sinh\sqrt{\nu}x)$,
and the corresponding energy is
\begin{eqnarray}
E_n=\frac{c^2}{4\nu}  - \left(n+\frac{1}{2}\right)^2\nu
-ic\left(n+\frac{1}{2}\right)~. \label{E-1}
\end{eqnarray}
Note that $E_n$ is independent of $d$, which is a general feature
of Scarf-type potentials. As in Case 5, the imaginary part is
proportional to $n+1/2$, which is characteristic of black hole
QNMs.  We mention here that there is another complexification
scheme for Case 1, giving a singular potential which exhibits very
peculiar features, such as the existence of continuous bound
states spectrum. As this is not related to QNM, it will be
reported elsewhere \cite{CH}.

Case 2 potential with QNM is
\begin{eqnarray}
V(x)&=&-\frac{cd}{2\nu} \coth(\sqrt{\nu} x){\rm
cosech}~(\sqrt{\nu} x)\nonumber\\ & -&\frac{1}{4\nu}\left(\nu^2
+c^2 + d^2\right){\rm cosech}^2 (\sqrt{\nu} x)~,
\end{eqnarray}
where $ x\in (0,\infty)$.  Its normalizable counterpart is the
generalized P\"oschl-Teller potential \cite{Cooper}. The wave
function is $\psi_n(x)=(\sinh\sqrt{\nu} x)^{(ic+\nu)/2\nu}(\tanh
(\sqrt{\nu}x/2))^{id/2\nu} \chi_n(\cosh\sqrt{\nu} x)$.  The
energies are exactly given by Eq.~(\ref{E-1}).  The two cases
share the same QNM spectrum, despite the difference in the forms
of the potentials and wave functions.  This is not surprising, as
it only reflects the same relation between the two counterpart
potentials, the Scarf II and the generalized P\"oschl-Teller
potential (see eg., Table~1 in \cite{Cooper}).

To summarize, we have demonstrated that it is possible to extend
the usual QES theory to accommodate QNM solutions, by
complexifying certain parameters defining the QES potentials. We
found that  the five $sl(2)$-based QES systems listed in
\cite{GKO} can be so extended.   While  one of these cases admits
QES QNM, the other four cases give exact QNM solutions. It is
hoped that our work would motivate the search of many more
exact/quasi-exact systems of QNM in QES theories based on higher
Lie algebras, and in higher dimensions.

\vskip 1.5 truecm

This work was supported in part by the National Science Council of
the Republic of China under the Grants NSC 94-2112-M-032-009
(H.T.C.) and NSC 94-2112-M-032-007 (C.L.H.)


\begin{references}
\bibitem{QNM} For a good review, see eg.:
K.D. Kokkotas and B.G. Schmidt, Living Rev. Rel. {\bf 2}, 2
(1999); S. Chandrasekhar, {\sl The mathematical theory of black
holes} (Clarendon, Oxford, 1983).
\bibitem{alpha} G.Gamow, Z. Phys. {\bf 51}, 204 (1928);
R.W. Curney and E.U. Condon, Phys. Rev. {\bf 33}, 127 (1929).
\bibitem{spin} H.T. Cho, Phys. Rev. D {\bf 68}, 024003 (2003);
J. Natario and R. Schiappa, Adv. Theor. Math. Phys. {\bf 8}, 1001
(2004); H.T. Cho, Phys. Rev. {\bf 73}, 024019 (2006).
\bibitem{Barton} G. Barton, Ann. Phys. {\bf 166}, 322
(1986).
\bibitem{Kim} S.P. Kim, J. Korean Phys. Soc. {\bf 49}, 764 (2006).
gr-qc/0512005 (2005).
\bibitem{FM} V. Ferrari and B. Mashhoon, Phys. Rev. Lett. {\bf
52}, 1361 (1984).
\bibitem{TU} A. Turbiner and A.G. Ushveridze, Phys. Lett. A {\bf 126}, 181
(1987).
\bibitem{Tur} A.V.Turbiner, Comm. Math. Phys. {\bf 118}, 467 (1988).
\bibitem{GKO} A. Gonz\'alez, N. Kamran and P.J. Olver, Comm. Math. Phys.
{\bf 153}, 117 (1993).
\bibitem{Ush} A.G. Ushveridze,  Sov. Phys.-Lebedev Inst. Rep. {\bf 2},
50, 54 (1988); {\sl Quasi-exactly solvable models in quantum
mechanics} (IOP, Bristol, 1994).
\bibitem{PT} G. Post and A. Turbiner, Russian J. Math. Phys. {\bf
3}, 113 (1995).
\bibitem{KMO} N. Kamran, R. Milson and P.J. Olver,
Invariant modules and the reduction of nonlinear partial
differential equations to dynamical systems (1999).
solv-int/9904014.
\bibitem{Rel}
Y. Brihaye and P. Kosinski, Mod. Phys. Lett. A {\bf 13}, 1445
(1998); C.-L. Ho and V.R. Khalilov, Phys. Rev. A {\bf 61}, 032104
(2000); C.L. Ho and P. Roy, J. Phys. A {\bf 36}, 4617 (2003); Ann.
Phys. {\bf 312}, 161 (2004); C.L. Ho, Ann. Phys. Ann. Phys. 321,
2170 (2006); Y. Brihaye and A. Nininahazwe, Mod. Phys. lett. A
{\bf 20}, 1875 (2005).
\bibitem{BB} C.M. Bender and and S. Boettcher, J. Phys. A {\bf 31},
L273 (1998).
\bibitem{other} See \cite{Ush} for the analytic approach,
\cite{PT} on classification of one-dimensional QES operators
possessing finite-dimensional invariant subspace with a basis of
monomials, and \cite{KMO} on formulation extending to nonlinear
operators.
\bibitem{Cooper}  F. Cooper, A. Khare and U. Sukhatme, Phys. Rep.
{\bf 251}, 267 (1995).
\bibitem{CH}  H.-T. Cho and C.-L. Ho, Continuous bound state spectra,
quasi-exact solvability, and total transmission modes, Tamkang
preprint (Jun 2006). quant-ph/0606144.
\end{references}
\end{document}